\documentclass[10pt,twocolumn,letterpaper]{article}

\usepackage{cvpr}              

\usepackage{graphicx}
\usepackage{amsmath}
\usepackage{amssymb}
\usepackage{booktabs}
\usepackage{amsfonts}
\usepackage{amssymb}
\usepackage{graphicx}

\usepackage[pagebackref,breaklinks,colorlinks]{hyperref}

\usepackage[capitalize]{cleveref}
\crefname{section}{Sec.}{Secs.}
\Crefname{section}{Section}{Sections}
\Crefname{table}{Table}{Tables}
\crefname{table}{Tab.}{Tabs.}

\def\confName{CVPR}
\def\confYear{2023}

\begin{document}

\title{VDPVE: VQA Dataset for Perceptual Video Enhancement }

               

\author{
Yixuan Gao$^{1}$$^{*}$ \and Yuqin Cao$^{1}$$^{*}$ \and Tengchuan Kou$^{1}$$^{*}$ \and Wei Sun$^{1}$ \and Yunlong Dong$^{2}$\and Xiaohong Liu$^{2}$$^{\dag}$
\and Xiongkuo Min$^{1}$$^{\dag}$\and Guangtao Zhai$^{1}$$^{\dag}$\\
Institute of Image Communication and Network Engineering, Shanghai Jiao Tong University$^{1}$\\
John Hopcroft Center, Shanghai Jiao Tong University$^{2}$\\
{
\tt\small 
\{gaoyixuan,caoyuqin,2213889087,sunguwei,dongyunlong,xiaohongliu,}\\
{\tt\small minxiongkuo,zhaiguangtao\}@sjtu.edu.cn
}
\thanks{Equal contribution.}
\thanks{Corresponding authors.}
}

\maketitle

\begin{abstract}
\vspace{-10pt}
Recently, many video enhancement methods have been proposed to improve video quality from different aspects such as color, brightness, contrast, and stability.
Therefore, how to evaluate the quality of the enhanced video in a way consistent with human visual perception is an important research topic.
However, most video quality assessment methods mainly calculate video quality by estimating the distortion degrees of videos from an overall perspective. Few researchers have specifically proposed a video quality assessment method for video enhancement, and there is also no comprehensive video quality assessment dataset available in public.
Therefore, we construct a Video quality assessment dataset for Perceptual Video Enhancement (VDPVE) in this paper.
The VDPVE has 1211 videos with different enhancements, which can be divided into three sub-datasets: the first sub-dataset has 600 videos with color, brightness, and contrast enhancements; the second sub-dataset has 310 videos with deblurring; and the third sub-dataset has 301 deshaked videos.
We invited 21 subjects (20 valid subjects) to rate all enhanced videos in the VDPVE. After normalizing and averaging the subjective opinion scores, the mean opinion score of each video can be obtained.
Furthermore, we split the VDPVE into a training set, a validation set, and a test set, and verify the performance of several state-of-the-art video quality assessment methods on the test set of the VDPVE.
\end{abstract}

\begin{figure*}[ht]
\vspace{-80pt}
\begin{minipage}[b]{1.0\linewidth}
  \centering
  \vspace{1.5cm}
  \centerline{\includegraphics[scale=0.55, trim=20 105 30 40, clip]{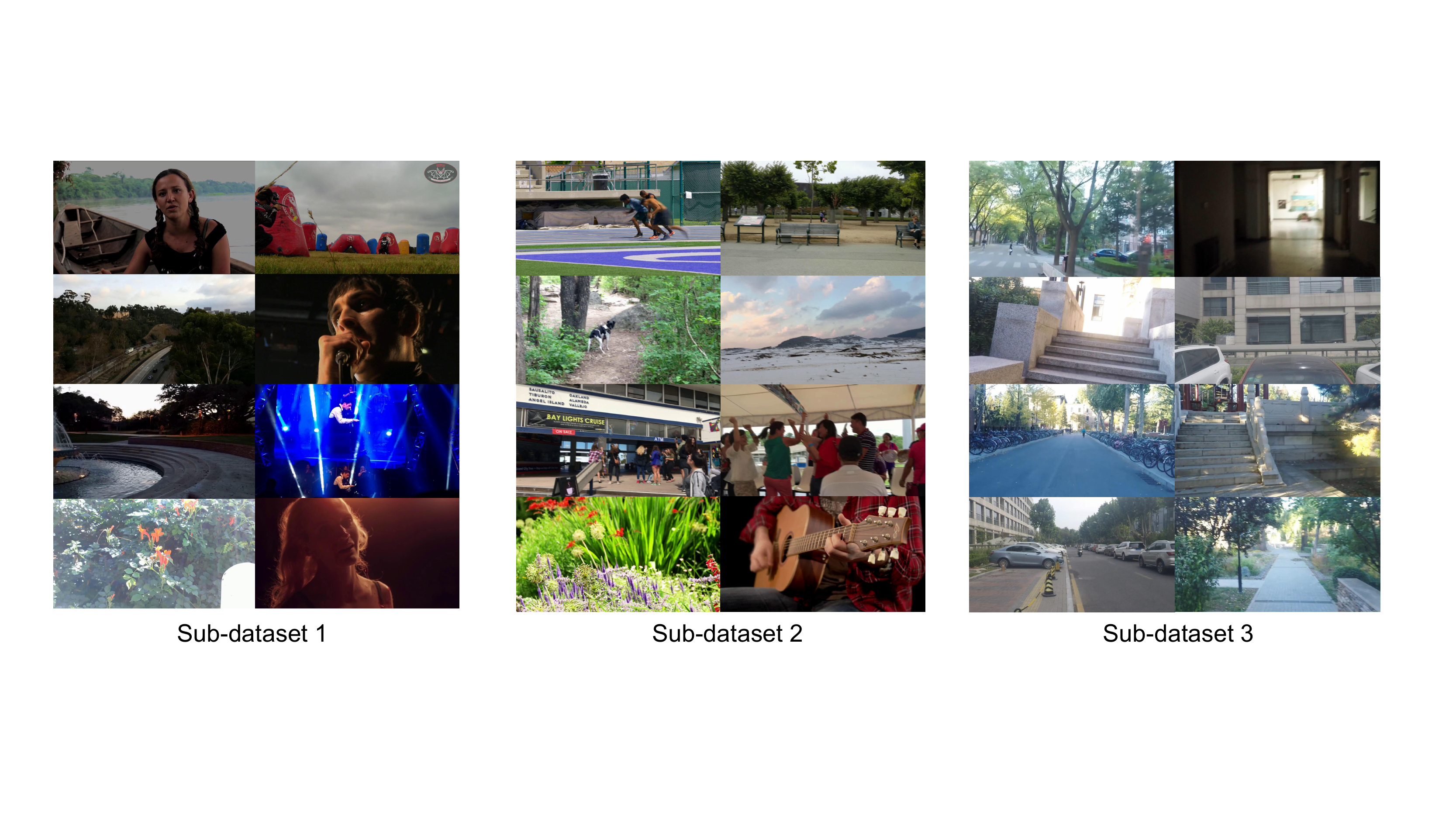}}
\end{minipage}
\vspace{-18pt}
\caption{Sample frames of representative enhanced videos in the three sub-datasets of VDPVE.
}
\label{ALL}
\end{figure*}
\vspace{-10pt}
\section{Introduction}
\label{sec:intro}
Due to the rapid development of social media applications, the demand for higher video quality is also increasing. Therefore, many video enhancement methods have emerged recently. These video enhancement methods aim to improve video quality from different aspects, such as color, contrast, brightness, and stability, to give people a more comfortable viewing experience. However, the quality of videos enhanced by different methods is distinct. Therefore, it is non-trivial to propose a promising method that enables the quality assessment of enhanced videos.

Traditional video quality assessment (VQA) methods assess the video quality mainly dependent on the distortions in video. Subjective VQA refers to rating the quality of videos by inviting many subjects to participate in a subjective experiment.
Objective VQA refers to calculating video quality by approximating the perceptions of people about video quality through various computational models.
For subjective VQA, researchers first process the obtained subjective opinion scores, such as screening invalid subjects and normalizing scores, and then average all the subjective opinion scores for each video to get the mean opinion score (MOS).
For example, the LIVE Video Quality Database \cite{seshadrinathan2010study}, proposed by Seshadrinanathan \emph{et al}. in 2010, is the most famous VQA dataset. This dataset contains 10 original videos with different contents and their corresponding distorted videos that are compressed and transmitted. A total of 38 subjects were invited to participate in the subjective experiment. In addition, other well-known VQA datasets include CSIQ \cite{vu2014vis}, MCL-V \cite{lin2015mcl}, and MCL-JCV \cite{wang2016mcl}. The distorted videos in these datasets are synthetically distorted and cannot represent videos with authentic distortions. Therefore, many user-generated content (UGC) VQA datasets were proposed. For example, the first UGC VQA dataset is the Camera Video Database (CVD2014) \cite{nuutinen2016cvd2014}. This dataset includes 234 authentically distorted videos captured by 78 different video capture devices. In addition, the authors in \cite{hosu2020konstanz} proposed the KoNViD-1k dataset. This dataset consists of 1200 videos, and 642 subjects were invited to rate these videos. The video quality obtained by subjective VQA is usually considered to be consistent with people's perceptions.

However, subjective VQA has some disadvantages, such as complex preparation, high cost, and time-consuming, which make it difficult to apply in practice. Objective VQA can avoid the disadvantages of subjective VQA, which can be specifically divided into full-reference (FR) VQA \cite{bampis2018spatiotemporal}, reduced-reference (RR) VQA \cite{soundararajan2012video}, and no-reference (NR) VQA \cite{saad2014blind}.
The FR VQA method requires all the information about the reference video to calculate the quality of the distorted video, for example, SSIM \cite{wang2004image} and ST-VMAF \cite{bampis2018spatiotemporal}.
The RR VQA method only requires partial information about the reference video to calculate the quality of the distorted video \cite{soundararajan2012video}. The NR VQA method, on the other hand, does not require any information about the reference video to calculate the quality of the distorted video. For example, the well-known V-BLIINDS \cite{saad2014blind} used a natural bandpass spatio-temporal video statistics model to calculate video quality. Kim \emph{et al}. \cite{kim2018deep} proposed a Deep Video Quality Assessor (DeepVQA), which predicts video quality by learning a spatiotemporal visual sensitivity map. Recently, Sun \emph{et al}.  \cite{sun2022deep} proposed a VQA method developed specifically for UGC videos. This method predicts the quality of UGC videos by training an end-to-end spatial feature extraction network to learn quality-aware spatial feature representations directly from the original pixels of a video frame.
Compared with subjective VQA, objective VQA is more convenient, faster, and easier to apply in the industry. However, objective VQA methods often require VQA datasets constructed by subjective VQA methods to verify their effectiveness and feasibility. Therefore, a successful VQA dataset can promote the development of objective VQA methods.

At present, few researchers have proposed objective VQA methods for video enhancement. The reason might be that there is no successful and effective VQA dataset for video enhancement. To break through this limitation, this paper constructs a VQA Dataset for Perceptual Video Enhancement (VDPVE) with a large number of videos and a wide range of enhancements. Specifically, the VDPVE includes 1211 enhanced videos, which can be divided into three sub-datasets: the first sub-dataset consists of 600 videos with color, brightness, and contrast enhancements; the second sub-dataset consists of 310 videos with deblurring; and the third sub-dataset consists of 301 deshaked videos.
We also carried out a subjective experiment and obtained 21 subjective opinion scores for each enhanced video.
During data processing, we reject an invalid subject and normalize the subjective opinion scores. Finally, we can obtain the mean opinion score (MOS) for each enhanced video.
The proposed VDPVE successfully fills in the blank of such kind of dataset in the field of video processing. It can not only promote the development of VQA methods for video enhancement, but also improve the performance of video enhancement methods to a certain extent.

The rest of the paper is organized as follows. In Section \ref{Dataset Construction}, details of the proposed VDPVE are provided. 
Section \ref{Video Quality Assessment} describes the subjective experiment in detail. 
In Section \ref{Experiment}, we test the performance of some VQA methods on the VDPVE.
Finally, we conclude this paper in Section \ref{Conclusion}.

\begin{figure*}[t]
\vspace{-40pt}
\begin{minipage}[b]{0.5\linewidth}
  \centering
  \vspace{1.5cm}
  \centerline{\includegraphics[scale=0.58, trim=80 180 415 40, clip]{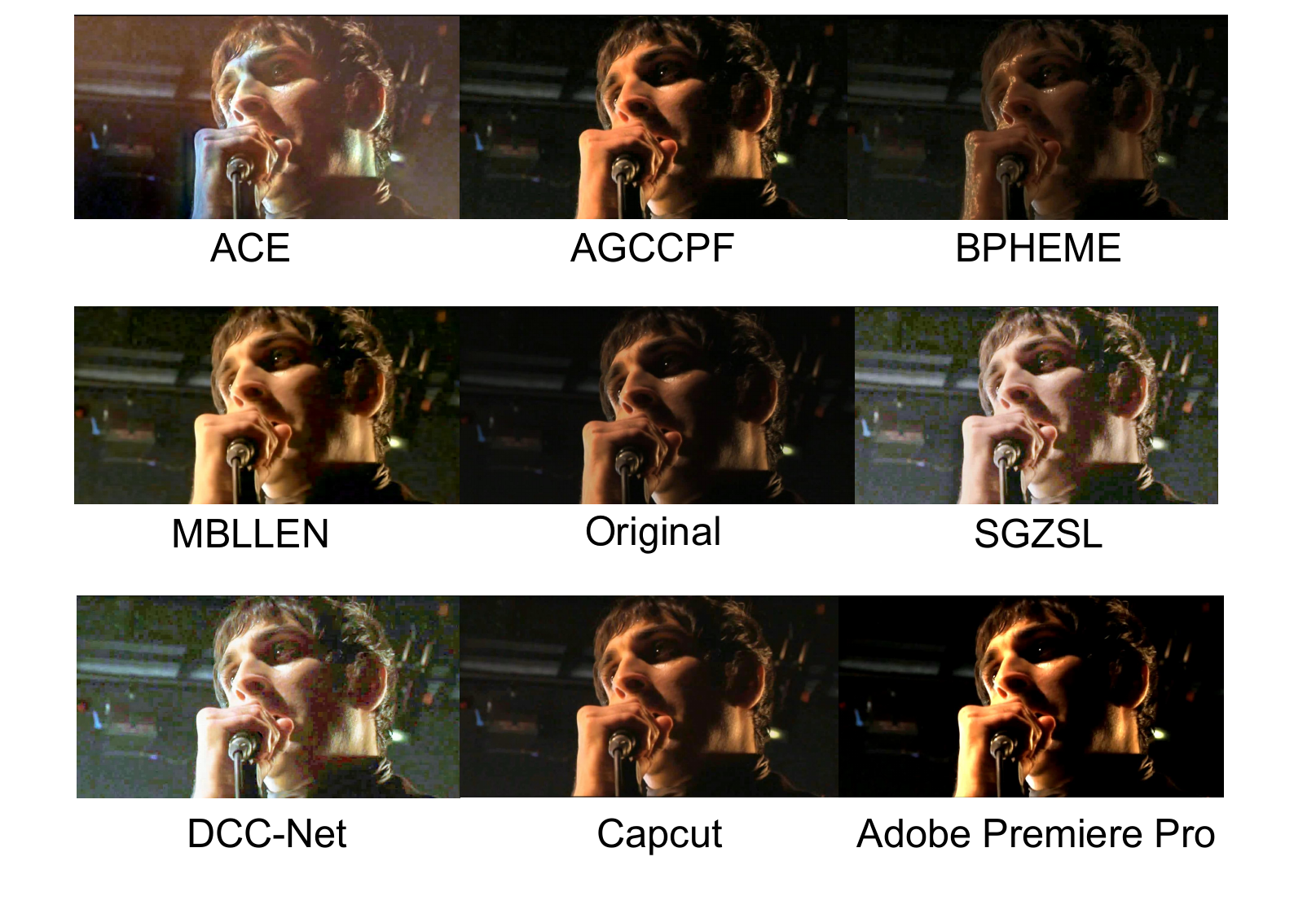}}
\end{minipage}
\begin{minipage}[b]{0.5\linewidth}
  \centering
  \vspace{1.5cm}
  \centerline{\includegraphics[scale=0.58, trim=80 180 415 40, clip]{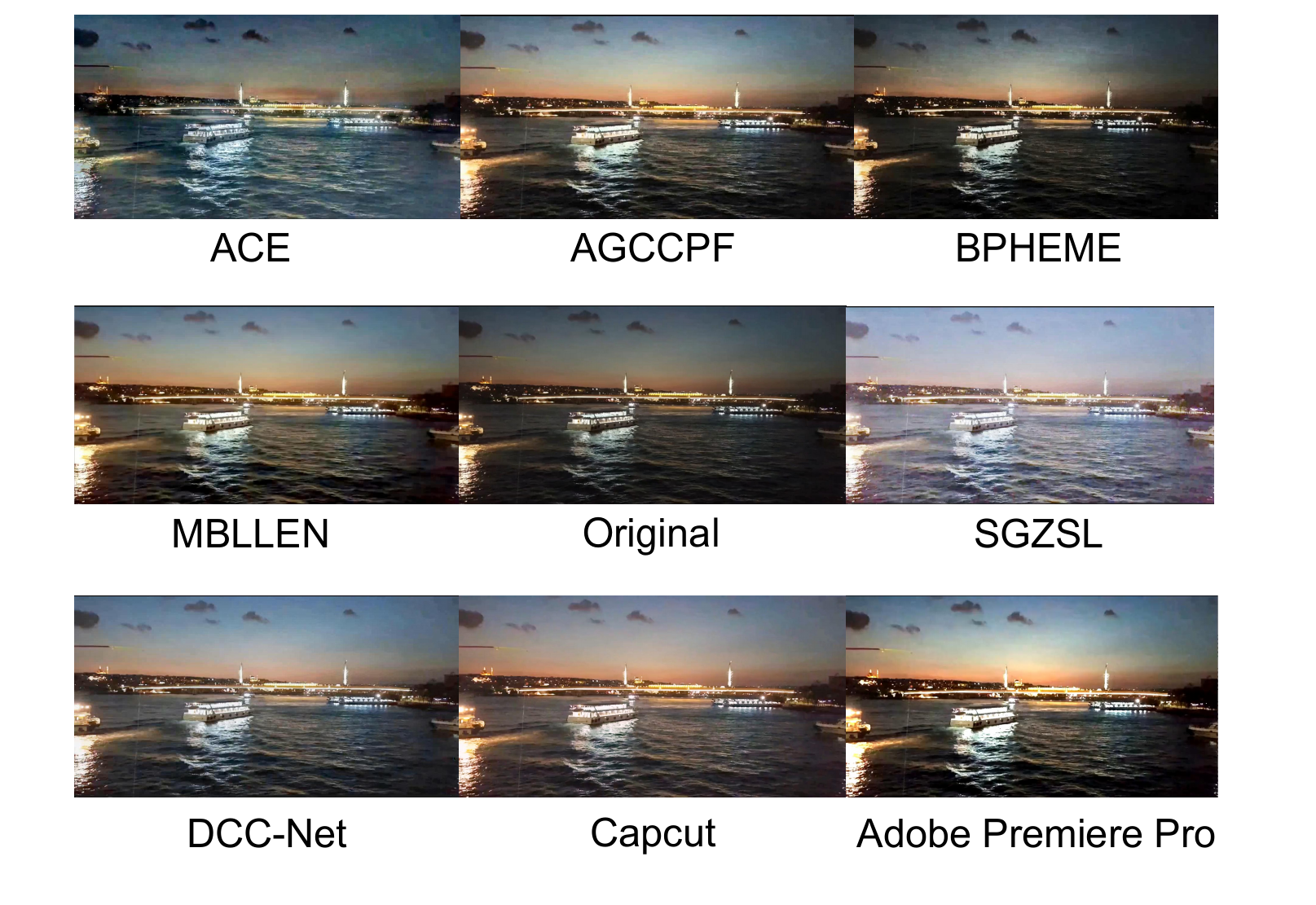}}
\end{minipage}
\vspace{-18pt}
\caption{Sample frames of two original videos and their corresponding enhanced videos in the sub-dataset 1.
}
\label{SD1}
\end{figure*}

\section{Dataset Construction}
\label{Dataset Construction}
 The VDPVE includes 1211 enhanced videos in total. Specifically, they can be divided into three sub-datasets.
The first sub-dataset consists of 600 videos with color, brightness, and contrast enhancements. The second sub-dataset consists of 310 videos with deblurring. The third sub-dataset consists of 301 deshaked videos.
Figure \ref{ALL} shows sample frames of representative videos in these three sub-datasets.
Next, we would like to introduce the construction process of these three sub-datasets in detail.

\subsection{Sub-dataset 1:}
The 600 videos with color, brightness, and contrast enhancements are obtained by using six enhancement methods on 79 original videos selected from four datasets.
\subsubsection{Original Video}
These 79 original videos are selected from the following four datasets:
	\begin{itemize}
\item  LIVE-Qualcomm \cite{ghadiyaram2017capture}: the dataset contains 208 videos in total. These videos have a duration of 15s and are subject to one of the following six distortions: artifacts, color, exposure, focus, sharpness, and stabilization. These videos were captured by the following eight mobile devices: Samsung Galaxy S5, Samsung Galaxy S6, HTC One VX, Apple iPhone 5S, Nokia Lumia 1020, LG G2, Samsung Galaxy Note4, and Oppo Find 7, which are widely used to capture videos. All videos have a resolution of 1920$\times$1080.
\item  V3C1 \cite{berns2019v3c1}: the dataset is the first partition of the Vimeo Creative Commons Collection (V3C) \cite{rossetto2019v3c}, which is designed to represent real network videos with diverse video contents. There are 7475 videos in the V3C1 dataset, which were selected from videos uploaded to Vimeo from 2006 to 2018. The resolutions of most videos are 1280$\times$720 and 1920$\times$1080.
\item  KoNViD-1k \cite{hosu2020konstanz}: the dataset is composed of 1200 videos, which were all obtained from the large public video dataset YFCC100m \cite{thomee2016yfcc100m} through “fair sampling”. The video length in KoNViD-1k is 8s, and the following three frame rates are used for encoding: 24 frames per second (FPS), 25 FPS, and 30 FPS, which correspond to 27$\%$, 5$\%$, and 68$\%$ of the videos, respectively. There are 12 resolutions in total, and the resolution of most videos is 1280$\times$720, followed by 1920$\times$1080.
\item  LIVE-VQC \cite{sinno2018large}: the dataset contains 585 videos with unique content that were captured by 101 different devices (most of them are smart phones) and has extensive authentic distortions. The main resolutions are 404$\times$720, 1024$\times$720, and 1920$\times$1080. The average video duration is 10s.
	\end{itemize}
 
We select 7, 26, 13, and 33 original videos from the LIVE-Qualcomm, V3C1, KoNViD-1k, and LIVE-VQC datasets, respectively. We set the resolution of all original videos to 1280$\times$720, and the duration is 8s or 10s.

\subsubsection{Enhancement Method}
We use eight enhancement methods to enhance the color, brightness, and contrast of these original videos, including
ACE \cite{getreuer2012automatic}, AGCCPF \cite{gupta2016minimum}, BPHEME \cite{wang2005brightness}, MBLLEN \cite{lv2018mbllen}, SGZSL \cite{zheng2022semantic}, DCC-Net \cite{zhang2022deep}, and two commercial softwares: CapCut and Adobe Premiere Pro.
Specifically, ACE \cite{getreuer2012automatic} is an effective color correction and enhancement method based on a simple model of the human visual system.
AGCCPF \cite{gupta2016minimum} uses the gamma correction of brightness pixels and weighted probability distribution to enhance the contrast and brightness of the image.
BPHEME \cite{wang2005brightness}, namely brightness preserving histogram equalization with maximum entropy, is a brightness enhancement method.
MBLLEN \cite{lv2018mbllen}, SGZSL \cite{zheng2022semantic}, and DCC-Net \cite{zhang2022deep} are low-light enhancement methods.
CapCut and Adobe Premiere Pro are popular video editing softwares. Users can enhance the color, contrast, and brightness of videos through these two softwares.

For ACE \cite{getreuer2012automatic}, AGCCPF \cite{gupta2016minimum}, BPHEME \cite{wang2005brightness}, MBLLEN \cite{lv2018mbllen}, SGZSL \cite{zheng2022semantic}, and DCC-Net \cite{zhang2022deep}, we use the default parameters to enhance videos. When using Adobe Premiere Pro to process videos, the brightness and contrast parameters are set to 50.

\subsubsection{Enhanced Video}
We enhance the color, brightness, and contrast of 79 original videos through the above eight enhancement methods, and obtain a total of 632 enhanced videos. We delete 32 videos with extremely serious distortions and finally obtain 600 videos with color, brightness, and contrast enhancements. 
Figure \ref{SD1} shows the sample frames of two original videos and their corresponding enhanced videos in the sub-dataset 1.

\begin{figure}[t]
\vspace{-40pt}
\begin{minipage}[b]{1.0\linewidth}
  \centering
  \vspace{1.5cm}
  \centerline{\includegraphics[scale=0.5, trim=250 110 250 140, clip]{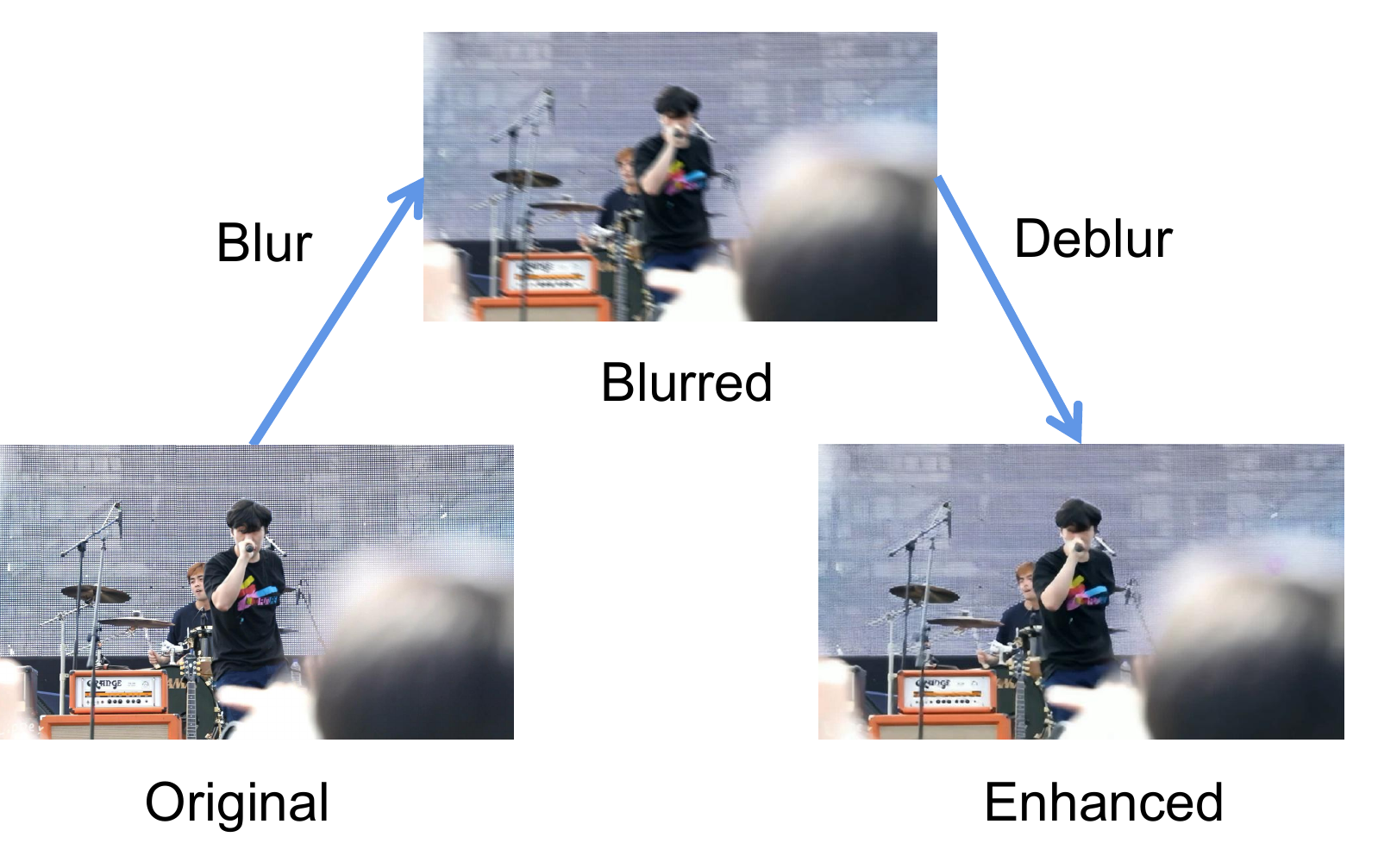}}
\end{minipage}
\vspace{-18pt}
\caption{Process of enhancing videos in the sub-dataset 2.
}
\label{blur}
\end{figure}
\subsection{Sub-dataset 2:}
The 310 videos with deblurring constitute the second sub-dataset of the VDPVE. 
These enhanced videos are obtained by using five deblurring methods on 62 original videos selected from four datasets.
\subsubsection{Original Video}
We first selected 22, 24, 10, and 6 videos from the LIVE-VQC \cite{sinno2018large}, Youtube-UGC \cite{wang2019youtube}, LIVE-YT-HFR \cite{madhusudana2021subjective}, and BVI-HFR \cite{mackin2018study} datasets, respectively.
	\begin{itemize}
\item Youtube-UGC \cite{wang2019youtube}: this dataset consists of 1098 20-second videos, which were sampled from YouTube. These videos are not always professionally curated and often suffer from diverse authentic distortions, like blockiness, blur, banding, noise, and so on.
The frame rates of videos vary from 15 to 60 FPS. 
\item LIVE-YT-HFR \cite{madhusudana2021subjective}: this dataset is comprised of 480 10-second videos with 6 different frame rates, obtained from 16 diverse contents. The videos are processed at five compression levels at each frame rate. The frame rates of videos vary from 24 to 120 FPS. 
\item BVI-HFR \cite{mackin2018study}: this dataset contains 88 videos with four different frame rates (from 15 to 120 FPS), including 22 120 FPS source sequences that were captured natively using a RED Epic-X video camera. 
	\end{itemize}
	We select 22 and 24 original videos with a frame rate of 30 FPS from the LIVE-VQC and Youtube-UGC datasets, and select 10 and 6 original videos with a frame rate of 120 FPS from the LIVE-YT-HFR and BVI-HFR datasets. We set the resolution of all original videos to 1280 $\times$ 720, and the duration is 8s.
	
	Post-processing is then performed on these original videos to obtain the blurred videos. On the one hand, we manually add motion blur to the original videos selected from the LIVE-VQC and Youtube-UGC datasets to obtain the blurred videos through the OpenCV toolbox.
	On the other hand, for the original videos selected from the LIVE-YT-HFR and BVI-HFR datasets, we take every 8 frames of each video as a group and average them as the current frame, where the adjacent groups overlap 4 frames, to convert the original videos with a frame rate of 120 FPS into the blurred videos with a frame rate of 30 FPS.
\begin{figure*}[t]
\vspace{-40pt}
\begin{minipage}[b]{0.5\linewidth}
  \centering
  \vspace{1.5cm}
  \centerline{\includegraphics[scale=0.60, trim=60 80 480 40, clip]{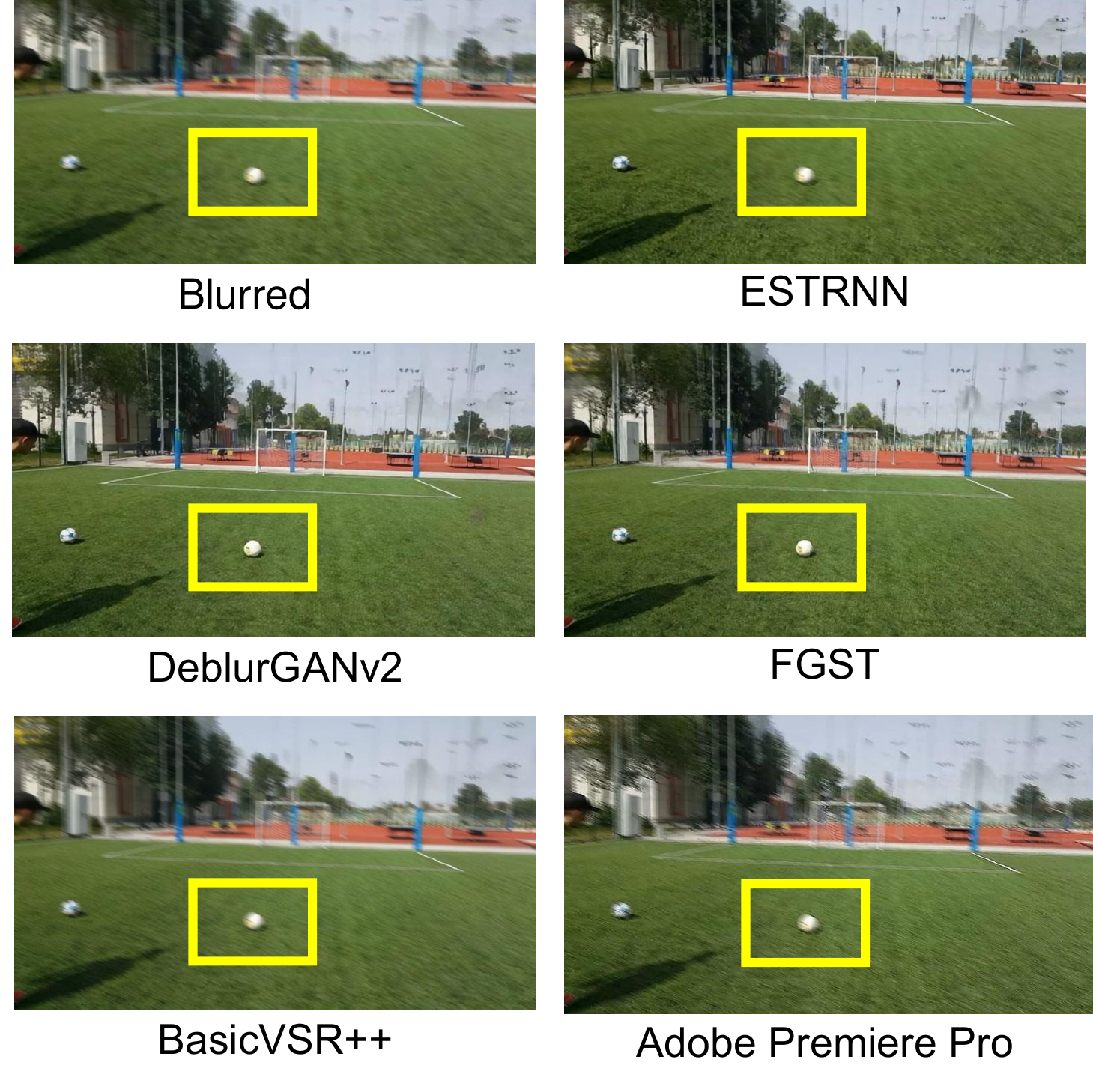}}
\end{minipage}
\begin{minipage}[b]{0.5\linewidth}
  \centering
  \vspace{1.5cm}
  \centerline{\includegraphics[scale=0.60, trim=60 80 480 40, clip]{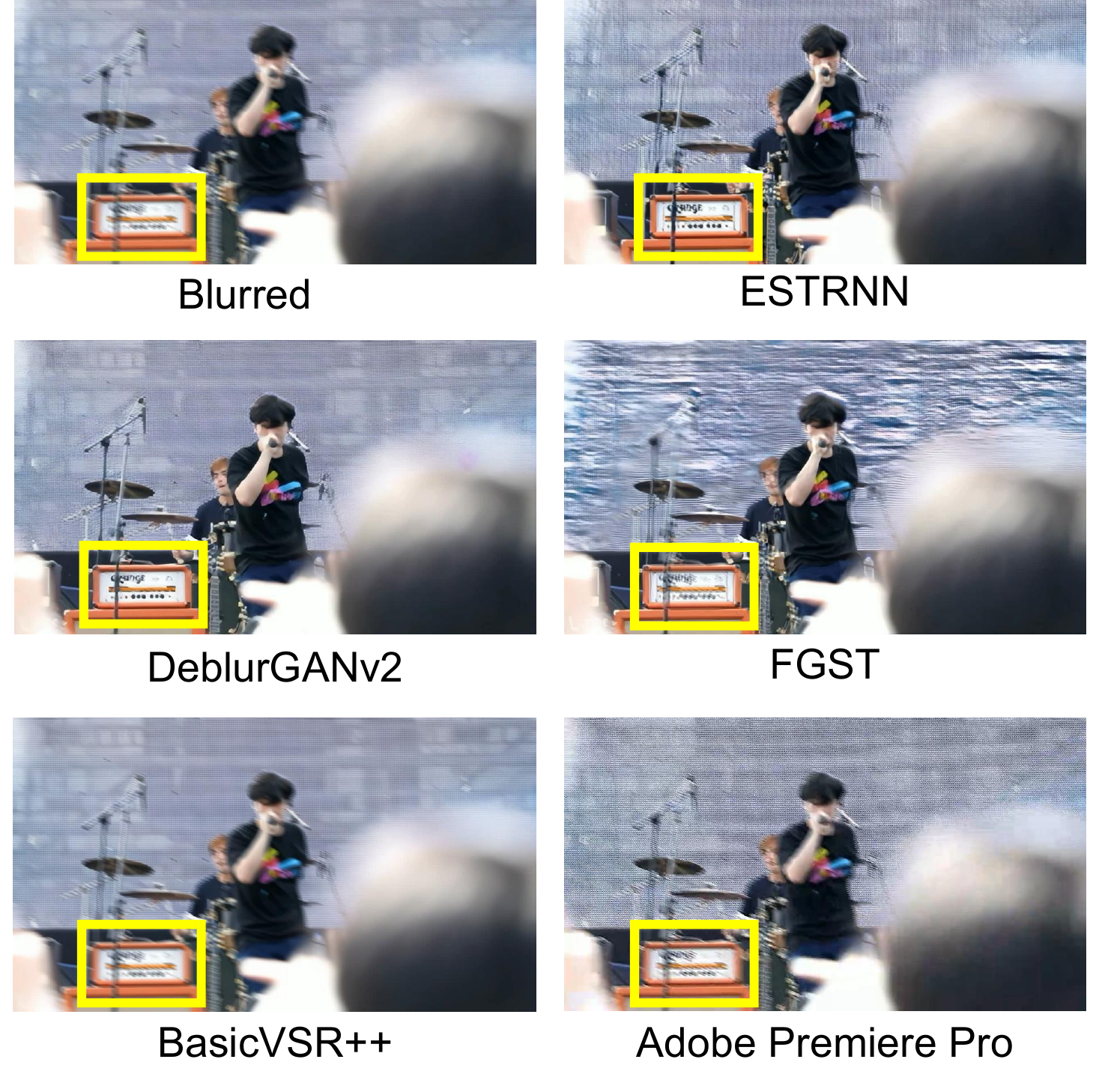}}
\end{minipage}
\vspace{-18pt}
\caption{Sample frames of two videos and their corresponding deblurred videos in the sub-dataset 2.
}
\label{SD2}
\end{figure*}

\subsubsection{Enhancement Method}
The following five enhancement methods are used to deblur the 62 blurred videos: ESTRNN \cite{zhong2020efficient}, DeblurGANv2 \cite{kupyn2019deblurgan}, FGST \cite{lin2022flow}, BasicVSR++\cite{chan2022basicvsr++}, and Adobe Premiere Pro.
Specifically, ESTRNN \cite{zhong2020efficient} realizes the deblurring of the current frame by fusing the effective layered features of the past and future frames.
DeblurGANv2 \cite{kupyn2019deblurgan} firstly introduces the feature pyramid network into deblurring, which can work flexibly with a variety of backbones to find a balance between performance and efficiency of deblurring.
FGST \cite{lin2022flow} proposes a flow-guided sparse window-based multi-head self-attention module to deblur videos.
BasicVSR++\cite{chan2022basicvsr++} proposes a second-order grid propagation and flow-guided deformable alignment model to effectively utilize the spatiotemporal information in unaligned video frames to achieve video deblurring.

For ESTRNN \cite{zhong2020efficient}, DeblurGANv2 \cite{kupyn2019deblurgan}, FGST \cite{lin2022flow}, and BasicVSR++\cite{chan2022basicvsr++}, we use the default parameters to enhance videos. When using Adobe Premiere Pro to process videos, we set the sharpness to 100.
\subsubsection{Enhanced Video}
Figure \ref{blur} shows the process of enhancing videos in the sub-dataset 2.
We deblur the 62 blurred videos through the above five enhancement methods, and obtain a total of 310 enhanced videos. 
Figure \ref{SD2} shows the sample frames of two blurred videos and their corresponding enhanced videos in the sub-dataset 2.

\begin{table*}[t]
\renewcommand\arraystretch{1.2}
\setlength{\tabcolsep}{4mm}
\begin{center}
\caption{Details of enhanced videos of all sub-datasets in the VDPVE. $\#$ means the number.
}
\label{t1}
\begin{tabular}{c||c|c|c|c|c}
  \hline\hline
Dataset &$\#$ Method& $\#$ Original &$\#$ Enhanced&Resolution&Duration  \\
  \hline\hline
Sub-dataset 1&8&79&600&1280 $\times$ 720&8s, 10s\\\hline
Sub-dataset 2&5&62&310&1280 $\times$ 720&8s\\\hline
Sub-dataset 3&7&43&301&1280 $\times$ 720&10s\\
\hline\hline
\end{tabular}
\end{center}
\vspace{-4pt}
\end{table*}

\begin{figure*}[t]
\begin{minipage}[(b) ]{0.5\linewidth}
  \centering
  \centerline{\includegraphics[scale=0.55, trim=100 250 100 250, clip]{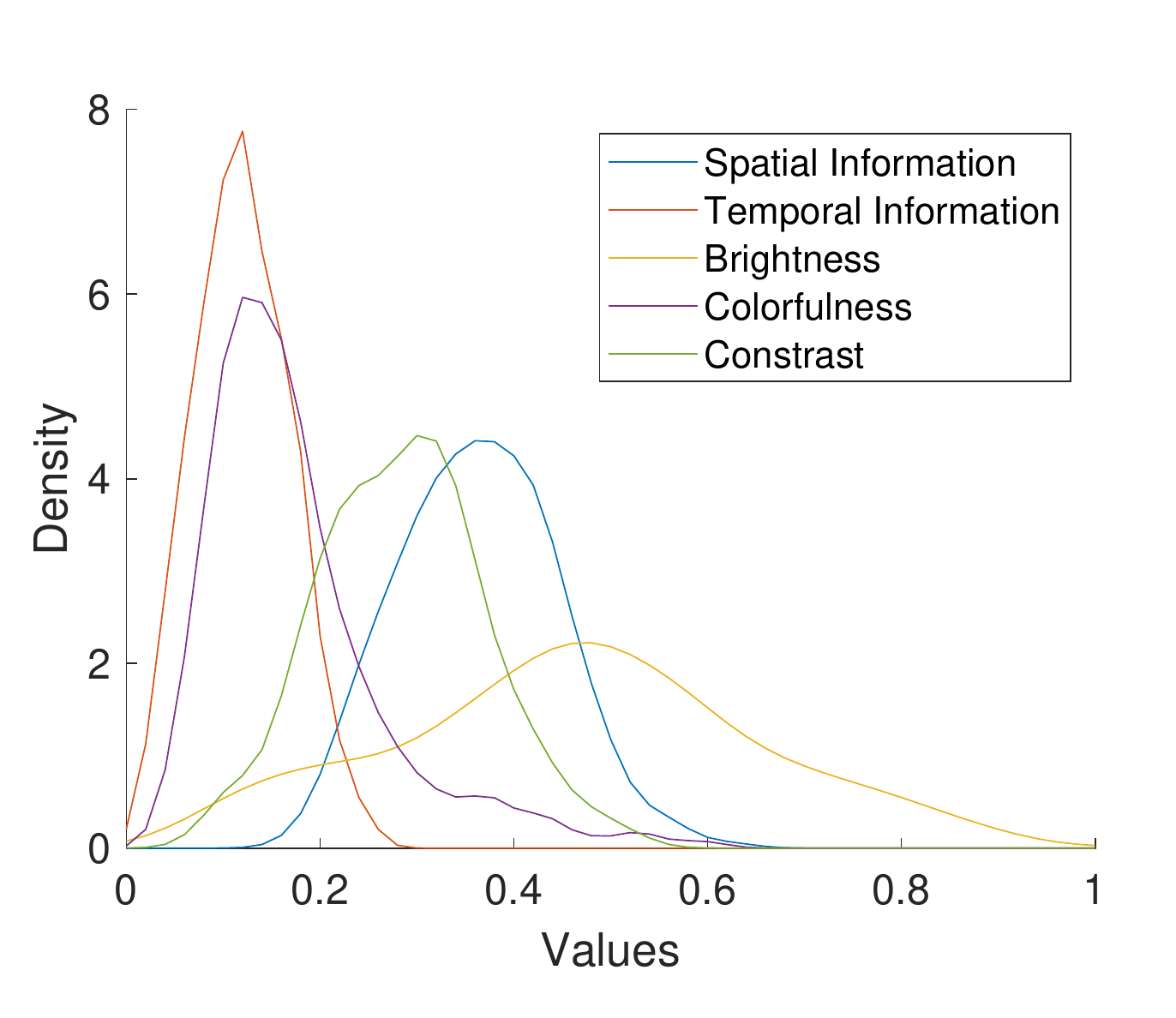}}
  \centerline{(a) Sub-dataset 1.}\medskip
  \end{minipage}
\begin{minipage}[(b) ]{0.5\linewidth}
  \centering
  \centerline{\includegraphics[scale=0.55,trim=100 250 100 250, clip]{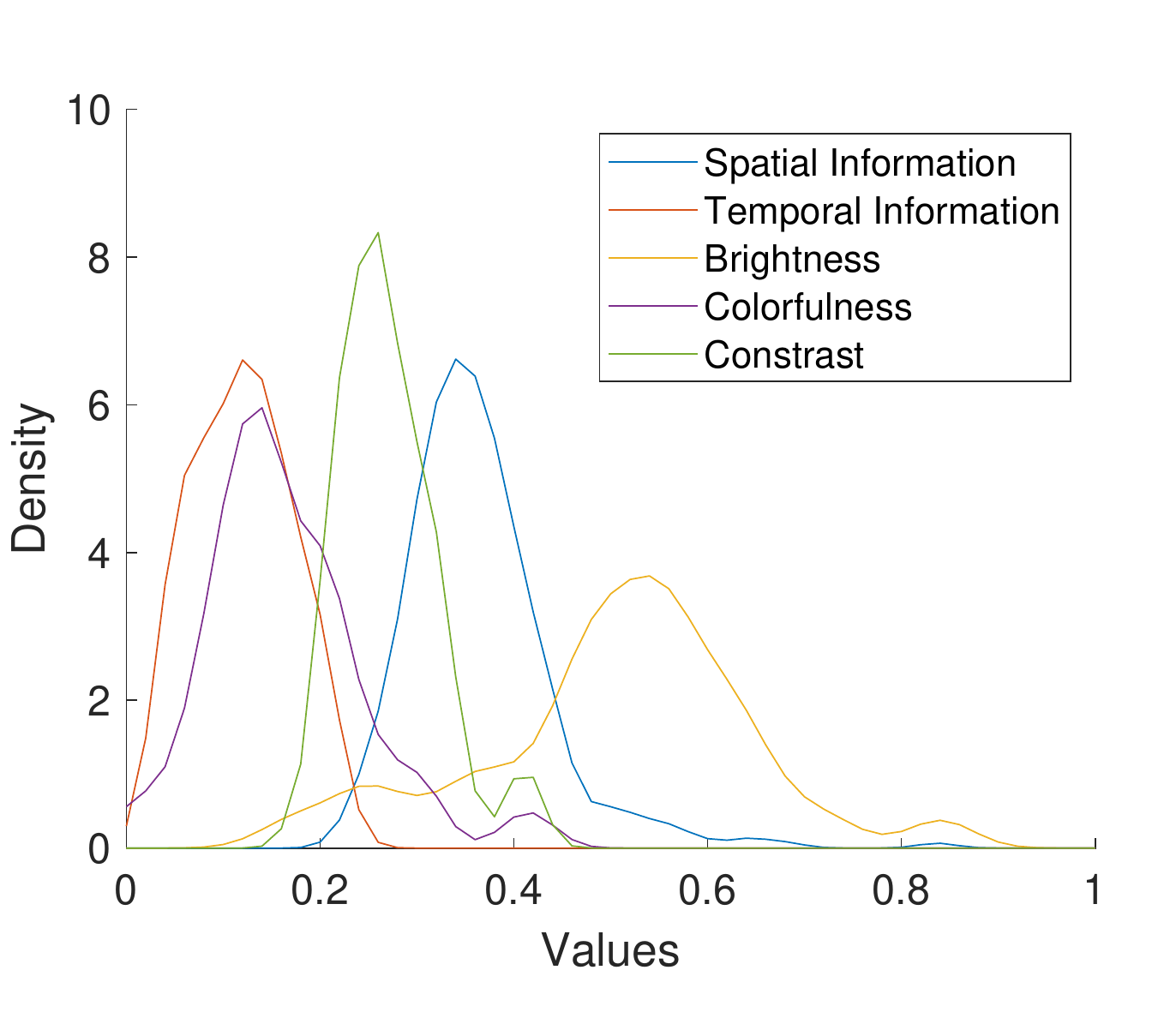}}
  \centerline{(b) Sub-dataset 2.}\medskip
\end{minipage}

\begin{minipage}[(b) ]{0.5\linewidth}
  \centering
  \centerline{\includegraphics[scale=0.55, trim=100 250 100 250, clip]{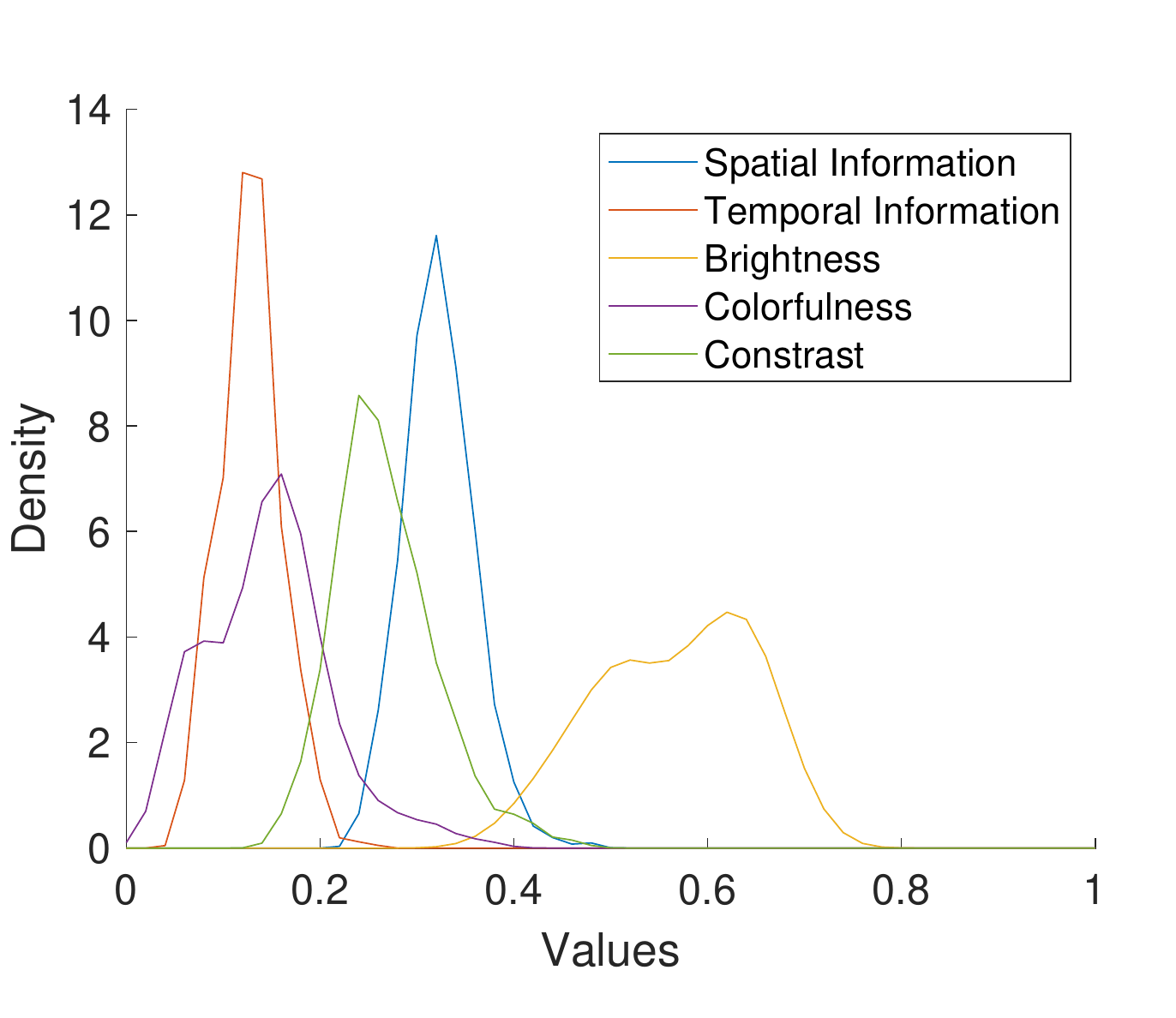}}
  \centerline{(c) Sub-dataset 3.}\medskip
  \end{minipage}
\begin{minipage}[(b) ]{0.5\linewidth}
  \centering
  \centerline{\includegraphics[scale=0.55,trim=100 250 100 250, clip]{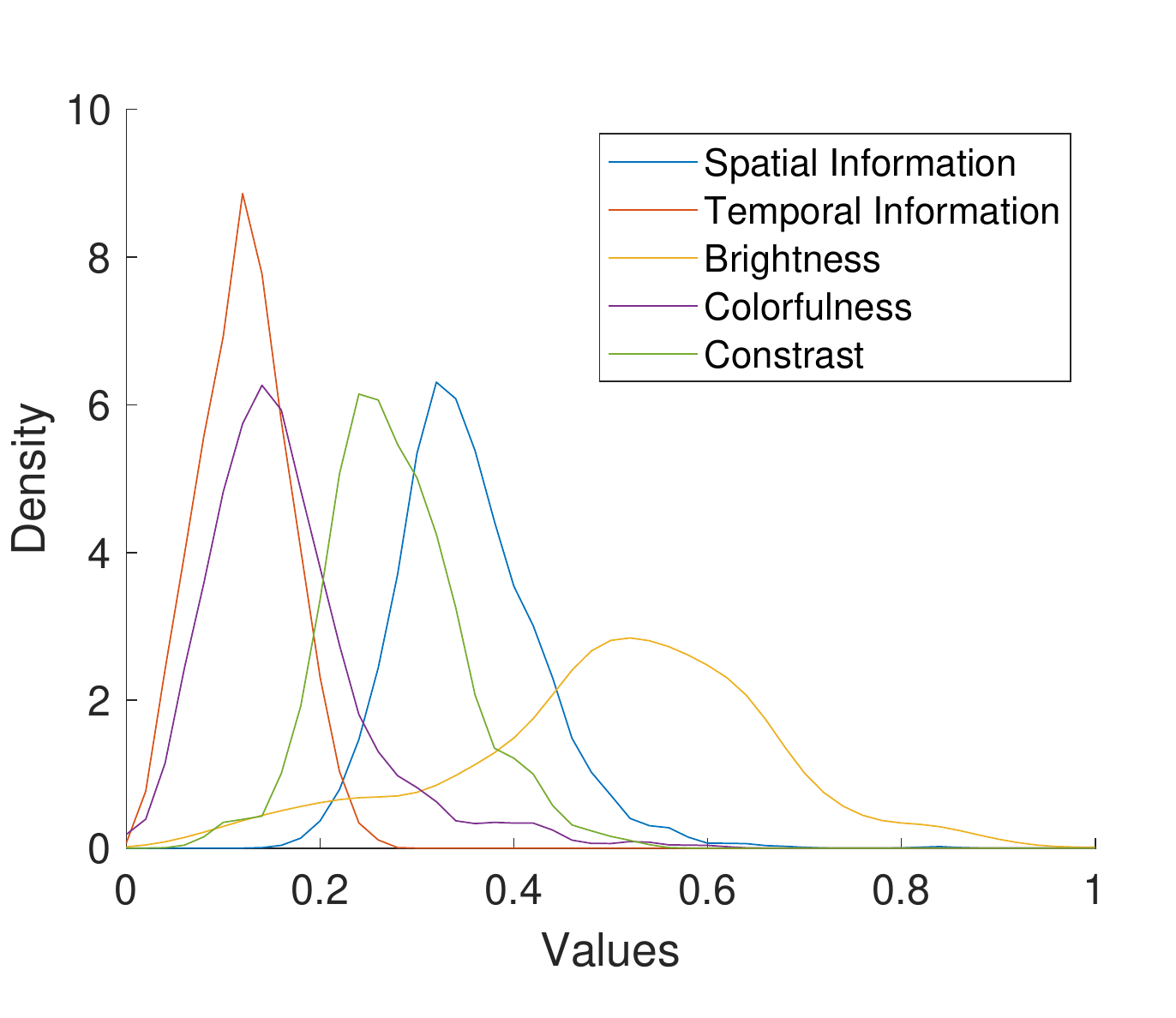}}
  \centerline{(d) VDPVE.}\medskip
\end{minipage}
\caption{Distribution of attribute values over the three sub-datasets and the VDPVE.
}
\vspace{-10pt}
  \label{feature}
\end{figure*}

\subsection{Sub-dataset 3:}
The 301 deshaked videos constitute the third sub-dataset of the VDPVE. 
These enhanced videos are obtained by using seven enhancement methods on 43 original videos selected from two datasets.
\subsubsection{Original Video}
We select 31 and 12 original videos from the DeepStab \cite{wang2018deep} and NUS \cite{liu2013bundled} datasets, respectively.

	\begin{itemize}
\item DeepStab \cite{wang2018deep}: this dataset contains 60 synchronized steady and unsteady video pairs captured by specially designed hand-held hardware.
The length of these videos is within 30s with a frame rate of 30 FPS.
\item NUS \cite{liu2013bundled}: this dataset contains 174 videos with a length of 10s to 60s. These videos were selected from previous publications, Internet, and the authors' own captures.
	\end{itemize} 
	 We set the resolution of all original videos to 1280$\times$720 and the duration to 10s.
\subsubsection{Enhancement Method}
We use seven enhancement methods to stabilize original videos, including GlobalFlowNet \cite{james2023globalflownet}, DIFRINT \cite{choi2020deep}, PWStableNet \cite{zhao2020pwstablenet}, Yu \cite{yu2020learning}, CapCut (most stable mode),  CapCut (minimum cropping mode), and Adobe Premiere Pro.
Specifically, GlobalFlowNet \cite{james2023globalflownet} uses a two-stage process to stabilize the video. 
DIFRINT \cite{choi2020deep} uses frame interpolation technology to generate in-between frames, thus reducing the shake between frames.
Zhao \emph{et al}. believed that different pixels might have different distortions, thus they proposed a pixel-level video stabilization method based on deep learning, PWStableNet \cite{zhao2020pwstablenet}. The proposed method is based on a multistage cascaded encoder-decoder architecture and learns per-pixel distortion mapping from continuous unstable frames.
Yu \cite{yu2020learning} uses optical flow to analyse motion and learn stability directly.

For GlobalFlowNet \cite{james2023globalflownet}, DIFRINT \cite{choi2020deep}, PWStableNet \cite{zhao2020pwstablenet}, and Yu \cite{yu2020learning}, we use the default parameters to enhance videos. When using CapCut to process videos, we use the most stable mode and the minimum cropping mode. When using Adobe Premiere Pro to process videos, we use the default parameters for stabilization.
\subsubsection{Enhanced Video}
Finally, we deshake 43 original videos through the above seven methods, and obtain a total of 301 enhanced videos. 
\subsection{Video Analysis}
Table \ref{t1} summarizes the details of enhanced videos in all sub-datasets in the VDPVE.
In order to measure the content diversity of the dataset, we calculate five video attributes of the VDPVE and three sub-datasets, including spatial information, temporary information, brightness, colorfulness, and contrast. Specifically, Figure \ref{feature}(a) shows the distribution of attribute values over the sub-dataset 1.
Figure \ref{feature}(b) shows the distribution of attribute values over the sub-dataset 2.
Figure \ref{feature}(c) shows the distribution of attribute values over the sub-dataset 3.
Figure \ref{feature}(d) shows the distribution of attribute values over the VDPVE.
\section{Video Quality Assessment}
\label{Video Quality Assessment}
In this section, we introduce the subjective experiment in detail.
\subsection{Subjects}
We invited 21 subjects from 20 to 30 years old to participate in this subjective experiment. Before the experiment, we tested the visual conditions of all subjects. All 21 subjects had normal (corrected) vision and color vision.

\subsection{Experimental Procedures}
The enhanced videos were displayed on the monitor at their original resolution. Subjects were asked to score the enhanced videos within the range of [0, 100]. The scoring criteria are as follows: a score in the range between 0 and 20 means that the quality of the enhanced video is poor; a score in the range between 20 and 40 means that the quality of the enhanced video is bad; a score in the range between 40 and 60 means that the quality of the enhanced video is fair; a score in the range between 60 and 80 means that that the quality of the enhanced video is good; a score in the range between 80 and 100 means that the quality of the enhanced video is excellent.

\subsection{Data Processing}
\begin{figure}[t]
\vspace{-40pt}
\begin{minipage}[b]{1.0\linewidth}
  \centering
  \vspace{1.5cm}
  \centerline{\includegraphics[scale=0.58, trim=95 260 120 280, clip]{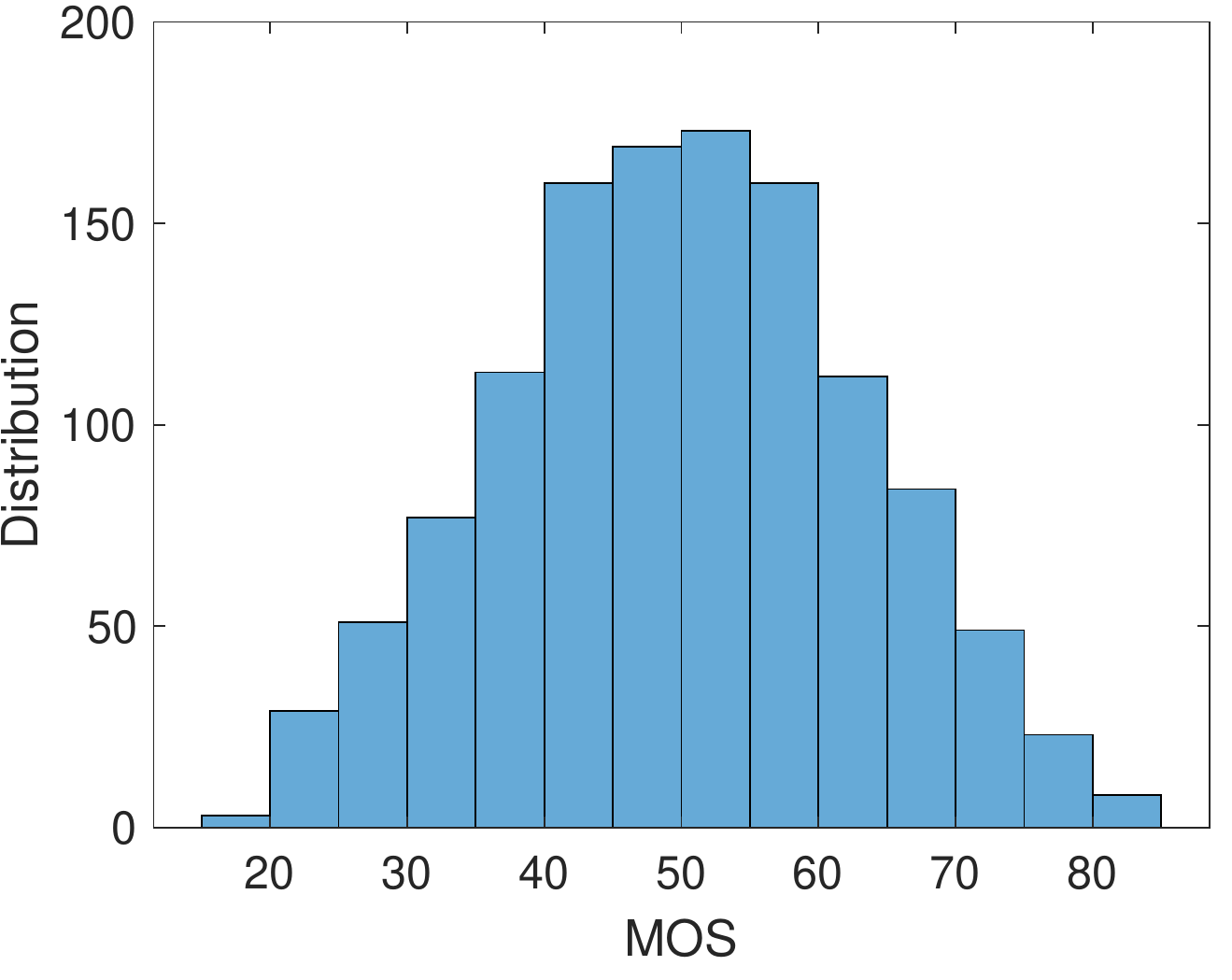}}
\end{minipage}
\vspace{-18pt}
\caption{Distribution of MOSs for all enhanced videos in the VDPVE.
}
\label{distribution}
\end{figure}
After the subjective experiment, each enhanced video has 21 subjective opinion scores. Then, we reject one invalid subject and obtain 20 subjective opinion scores for each video.
 Finally, we normalize the scores and calculate the MOS for each enhanced video.
Figure \ref{distribution} shows the distribution of MOSs for all enhanced videos in the VDPVE. It can be seen from the figure that the quality of enhanced videos in VDPVE is between 15 and 85.
We also calculate the mean 95$\%$ confidence intervals for different MOS ranges in Table \ref{t2}. It can be seen from the table that the length of the mean confidence interval of each MOS range will not exceed 10.

\begin{table}[t]
 \setlength{\tabcolsep}{9pt}
  \renewcommand{\arraystretch}{1.3}
\begin{center}
\caption{Mean confidence intervals for different MOS ranges. “CI” means the confidence interval.
}
\label{t2}
\begin{tabular}{c||c|c||c}
  \hline\hline
MOS & Mean CI&MOS & Mean CI  \\
  \hline\hline
[15, 20]&[12.6, 19.8]&[20, 25]&[19.1, 26.7]\\\hline
[25, 30]&[23.8,	32.4]&[30, 35]&[28.3, 36.8]\\\hline
[35, 40]&[32.9,	42.5]&[40, 45]&[37.6,	47.6]\\\hline
[45, 50]&[42.5,	52.0]&[50, 55]&[47.5,	57.4]\\\hline
[55, 60]&[52.5,	62.2]&[60, 65]&[57.6,	67.2]\\\hline
[65, 70]&[62.9,	71.7]&[70, 75]&[67.3,	77.0]\\\hline
[75, 80]&[72.9,	81.2]&[80, 85]&[78.0,	84.5]\\
\hline\hline
\end{tabular}
\end{center}
\vspace{-4pt}
\end{table}

\section{Experiment}
\label{Experiment}
In this section, some experiments are carried out on the proposed VDPVE.
\subsection{Set Splitting}
We first randomly split the enhanced videos in the VDPVE into a training set, a validation set, and a test set according to the ratio of 7:1:2. The enhanced videos generated from the same original video are divided into the same set, and the ratio of the training set, the validation set, and the test set of each divided sub-dataset is also about 7:1:2. This splitting process is repeated five times. Then, we test the performance of the latest VQA method, SimpleVQA \cite{sun2022deep}, on the validation sets and the test sets of the five splits.
We use the Pearson linear correlation coefficient (PLCC), Spearman rank correlation coefficient (SRCC), and root mean square error (RMSE) to measure the prediction performance of SimpleVQA \cite{sun2022deep} on the VDPVE with different splits, and report the results in Table \ref{split}.

From the table, we can see that SimpleVQA \cite{sun2022deep} has the most stable performance on the validation set and test set of the third split. Therefore, we finally split the VDPVE into a training set, a validation set, and a test set according to the third split. The numbers of enhanced videos in the training set, validation set, and test set are 839, 119, and 253, respectively.
\begin{table*}[t]
\renewcommand\arraystretch{1.2}
\setlength{\tabcolsep}{5mm}
\begin{center}
\caption{Performance of the latest VQA algorithm, SimpleVQA \cite{sun2022deep}, on the validation sets and the test sets of five splits.
}
\label{split}
\begin{tabular}{c||c|c|c|c|c|c}
  \hline\hline
  &\multicolumn{3}{c|}{Validation}   &\multicolumn{3}{c}{Test}\\ \hline
Split&	SRCC&	PLCC&	RMSE&	SRCC&	PLCC&	RMSE  \\
  \hline\hline
1&	0.7175&	0.6933	&8.4779&	0.7627&	0.7547&	8.9562\\\hline
2&	0.6992&	0.7161&	8.2571&	0.7435&	0.7635&	8.8168\\\hline
3&\textbf{0.6192}&	\textbf{0.6269}&	\textbf{10.1208}&	\textbf{0.6340}	&\textbf{0.6354}	&\textbf{9.8541}\\\hline
4&	0.7036&	0.7069&	8.6180	&0.7384	&0.7345&	9.9273\\\hline
5&	0.6624&	0.4810&	9.6621	&0.5658&	0.5545&	10.6574\\
\hline\hline
\end{tabular}
\end{center}
\vspace{-4pt}
\end{table*}

\begin{table*}[t]
\renewcommand\arraystretch{1.2}
\setlength{\tabcolsep}{7mm}
\begin{center}
\caption{Prediction performance of the state-of-the-art VQA methods on the test sets. “All” means the entire test set of the VDPVE. The best performances are in bold.
}
\label{test}
\begin{tabular}{c||c|c|c|c}
  \hline\hline
 Test Set
 &\multicolumn{2}{c|}{Sub-dataset 1}   &\multicolumn{2}{c}{Sub-dataset 2}\\ \hline
Method&	SRCC&	PLCC&	SRCC&	PLCC  \\
  \hline\hline
V-BLIINDS \cite{saad2014blind}&0.4561&	0.4959&	0.5048&	0.5408	\\\hline
TLVQM \cite{korhonen2019two}&0.5203	&0.5721	&0.2743	&0.3130\\\hline
VIDEVAL \cite{tu2021ugc}&0.3991&0.3972&	0.2887	&0.4348\\\hline
RAPIQUE \cite{tu2021rapique}&0.5161&0.5826&0.4711&	0.4789
\\\hline
FastVQA \cite{wu2022fast}&\textbf{0.7083}&	\textbf{0.7571}&	\textbf{0.7479}	&\textbf{0.7429}\\\hline
VSFA \cite{li2019quality}& 0.5240 &	0.4900 &	0.6196 &	0.6060 \\\hline
BVQA \cite{li2022blindly}&0.5742 &	0.5632 &	0.6025 &	0.6621 \\
\hline\hline
 Test Set
 &\multicolumn{2}{c|}{Sub-dataset 3}   &\multicolumn{2}{c}{All}\\ \hline
Method&	SRCC&	PLCC&	SRCC&	PLCC \\
  \hline\hline
V-BLIINDS \cite{saad2014blind}&0.6895&	0.6822&		0.5652&		0.5503\\\hline
TLVQM \cite{korhonen2019two}&	\textbf{0.8074}&		\textbf{0.8203}&		0.5474&		0.5509\\\hline
VIDEVAL \cite{tu2021ugc}&0.7111&0.6903	&0.5005&0.4724\\\hline
RAPIQUE \cite{tu2021rapique}&0.5554&	0.5736&0.5434&0.5393\\\hline
FastVQA \cite{wu2022fast}&	0.6368&		0.6433&	\textbf{0.7350}&		\textbf{0.7310}\\\hline
VSFA \cite{li2019quality}&	0.5874&	 	0.6281 	&	0.5871 &		0.5424 \\\hline
BVQA \cite{li2022blindly}&	0.7486 &	0.7671 &		0.6995&	 	0.6674 \\
\hline\hline

\end{tabular}
\end{center}
\vspace{-4pt}
\end{table*}

\subsection{Performance}
Then, we test the performance of several state-of-the-art VQA methods on the test set, including V-BLIINDS \cite{saad2014blind}, TLVQM \cite{korhonen2019two}, VIDEVAL \cite{tu2021ugc}, RAPIQUE \cite{tu2021rapique}, FastVQA \cite{wu2022fast}, VSFA \cite{li2019quality}, and BVQA \cite{li2022blindly}. Specifically, all methods are trained on the entire training set, verified on the entire validation set, and then tested on the three test sets of the three sub-datasets and the entire test set of the VDPVE, respectively.
PLCC and SRCC are used to measure the prediction performance of these state-of-the-art VQA methods.
Table \ref{test} shows the prediction performance of these state-of-the-art VQA methods. From the table, we can see that FastVQA \cite{wu2022fast} gets the best prediction performance on the sub-dataset 1, sub-dataset 2, and the entire VDPVE. The prediction performance of TLVQM \cite{korhonen2019two} in the sub-dataset 3 is the best.

\section{Conclusion}
\label{Conclusion}
To make up for the lack of the VQA dataset for video enhancement in the field of video processing, this paper constructs the VDPVE. The VDPVE includes a total of 1211 enhanced videos, involving a wide range of enhancement methods. Among them, there are 600 videos with color, contrast, and brightness enhancements; 310 videos with deblurring; and 301 deshaked videos. At the same time, we invited 21 subjects (20 valid subjects) to rate all enhanced videos in the VDPVE to obtain the MOS for each enhanced video. Finally, we split the VDPVE and verify the performance of several popular VQA methods on the test sets. The proposed VDPVE can provide the basis for the development of VQA methods for video enhancement, and further promote the performance of video enhancement methods.


{\small
\bibliographystyle{ieee_fullname}
\bibliography{egbib}

\begin{thebibliography}{10}\itemsep=-1pt

\bibitem{bampis2018spatiotemporal}
Christos~G Bampis, Zhi Li, and Alan~C Bovik.
\newblock Spatiotemporal feature integration and model fusion for full
  reference video quality assessment.
\newblock {\em IEEE Trans. Circuits Syst. Video Technol.}, 29(8):2256--2270,
  2018.

\bibitem{berns2019v3c1}
Fabian Berns, Luca Rossetto, Klaus Schoeffmann, Christian Beecks, and George
  Awad.
\newblock V3c1 dataset: an evaluation of content characteristics.
\newblock In {\em Proc. ACM Int. Conf. Multimed. Retr.}, pages 334--338, 2019.

\bibitem{chan2022basicvsr++}
Kelvin~CK Chan, Shangchen Zhou, Xiangyu Xu, and Chen~Change Loy.
\newblock Basicvsr++: Improving video super-resolution with enhanced
  propagation and alignment.
\newblock In {\em Proc. IEEE Comput. Soc. Conf. Comput. Vision Pattern
  Recognit.}, pages 5972--5981, 2022.

\bibitem{choi2020deep}
Jinsoo Choi and In~So Kweon.
\newblock Deep iterative frame interpolation for full-frame video
  stabilization.
\newblock {\em ACM Trans. Graph.}, 39(1):1--9, 2020.

\bibitem{getreuer2012automatic}
Pascal Getreuer.
\newblock Automatic color enhancement (ace) and its fast implementation.
\newblock {\em Image Process. On Line}, 2:266--277, 2012.

\bibitem{ghadiyaram2017capture}
Deepti Ghadiyaram, Janice Pan, Alan~C Bovik, Anush~Krishna Moorthy, Prasanjit
  Panda, and Kai-Chieh Yang.
\newblock In-capture mobile video distortions: A study of subjective behavior
  and objective algorithms.
\newblock {\em IEEE Trans. Circuits Syst. Video Technol.}, 28(9):2061--2077,
  2017.

\bibitem{gupta2016minimum}
Bhupendra Gupta and Mayank Tiwari.
\newblock Minimum mean brightness error contrast enhancement of color images
  using adaptive gamma correction with color preserving framework.
\newblock {\em Optik}, 127(4):1671--1676, 2016.

\bibitem{hosu2020konstanz}
V Hosu, F Hahn, M Jenadeleh, H Lin, H Men, T Szir{\'a}nyi, S Li, and D Saupe.
\newblock The konstanz natural video database, 2020.

\bibitem{james2023globalflownet}
Jerin~Geo James, Devansh Jain, and Ajit Rajwade.
\newblock Globalflownet: Video stabilization using deep distilled global motion
  estimates.
\newblock In {\em Proc. IEEE/CVF Winter Conf. Appl. Comput. Vis.}, pages
  5078--5087, 2023.

\bibitem{kim2018deep}
Woojae Kim, Jongyoo Kim, Sewoong Ahn, Jinwoo Kim, and Sanghoon Lee.
\newblock Deep video quality assessor: From spatio-temporal visual sensitivity
  to a convolutional neural aggregation network.
\newblock In {\em Proc. Eur. Conf. Comput. Vis.}, pages 219--234, 2018.

\bibitem{korhonen2019two}
Jari Korhonen.
\newblock Two-level approach for no-reference consumer video quality
  assessment.
\newblock {\em IEEE Trans. Image Process.}, 28(12):5923--5938, 2019.

\bibitem{kupyn2019deblurgan}
Orest Kupyn, Tetiana Martyniuk, Junru Wu, and Zhangyang Wang.
\newblock Deblurgan-v2: Deblurring (orders-of-magnitude) faster and better.
\newblock In {\em Proc. IEEE Int. Conf. Comput. Vision}, pages 8878--8887,
  2019.

\bibitem{li2022blindly}
Bowen Li, Weixia Zhang, Meng Tian, Guangtao Zhai, and Xianpei Wang.
\newblock Blindly assess quality of in-the-wild videos via quality-aware
  pre-training and motion perception.
\newblock {\em IEEE Trans. Circuits Syst. Video Technol.}, 32(9):5944--5958,
  2022.

\bibitem{li2019quality}
Dingquan Li, Tingting Jiang, and Ming Jiang.
\newblock Quality assessment of in-the-wild videos.
\newblock In {\em Proc. ACM Int. Conf. Multimedia}, pages 2351--2359, 2019.

\bibitem{lin2022flow}
Jing Lin, Yuanhao Cai, Xiaowan Hu, Haoqian Wang, Youliang Yan, Xueyi Zou,
  Henghui Ding, Yulun Zhang, Radu Timofte, and Luc Van~Gool.
\newblock Flow-guided sparse transformer for video deblurring.
\newblock {\em arXiv preprint arXiv:2201.01893}, 2022.

\bibitem{lin2015mcl}
Joe~Yuchieh Lin, Rui Song, Chi-Hao Wu, TsungJung Liu, Haiqiang Wang, and
  C-C~Jay Kuo.
\newblock {MCL-V}: A streaming video quality assessment database.
\newblock {\em J. Vis. Commun. Image Represent.}, 30:1--9, 2015.

\bibitem{liu2013bundled}
Shuaicheng Liu, Lu Yuan, Ping Tan, and Jian Sun.
\newblock Bundled camera paths for video stabilization.
\newblock {\em ACM Trans. Graph.}, 32(4):1--10, 2013.

\bibitem{lv2018mbllen}
Feifan Lv, Feng Lu, Jianhua Wu, and Chongsoon Lim.
\newblock Mbllen: Low-light image/video enhancement using cnns.
\newblock In {\em BMVC}, volume 220, page~4, 2018.

\bibitem{mackin2018study}
Alex Mackin, Fan Zhang, and David~R Bull.
\newblock A study of high frame rate video formats.
\newblock {\em IEEE Trans. Multimedia}, 21(6):1499--1512, 2018.

\bibitem{madhusudana2021subjective}
Pavan~C Madhusudana, Xiangxu Yu, Neil Birkbeck, Yilin Wang, Balu Adsumilli, and
  Alan~C Bovik.
\newblock Subjective and objective quality assessment of high frame rate
  videos.
\newblock {\em IEEE Access}, 9:108069--108082, 2021.

\bibitem{nuutinen2016cvd2014}
Mikko Nuutinen, Toni Virtanen, Mikko Vaahteranoksa, Tero Vuori, Pirkko
  Oittinen, and Jukka H{\"a}kkinen.
\newblock {CVD2014—A} database for evaluating no-reference video quality
  assessment algorithms.
\newblock {\em IEEE Trans. Image Process.}, 25(7):3073--3086, 2016.

\bibitem{rossetto2019v3c}
Luca Rossetto, Heiko Schuldt, George Awad, and Asad~A Butt.
\newblock V3c--a research video collection.
\newblock In {\em Proc. MultiMedia Model.}, pages 349--360. Springer, 2019.

\bibitem{saad2014blind}
Michele~A Saad, Alan~C Bovik, and Christophe Charrier.
\newblock Blind prediction of natural video quality.
\newblock {\em IEEE Trans. Image Process.}, 23(3):1352--1365, 2014.

\bibitem{seshadrinathan2010study}
Kalpana Seshadrinathan, Rajiv Soundararajan, Alan~Conrad Bovik, and Lawrence~K
  Cormack.
\newblock Study of subjective and objective quality assessment of video.
\newblock {\em IEEE Trans. Image Process.}, 19(6):1427--1441, 2010.

\bibitem{sinno2018large}
Zeina Sinno and Alan~Conrad Bovik.
\newblock Large-scale study of perceptual video quality.
\newblock {\em IEEE Trans. Image Process.}, 28(2):612--627, 2018.

\bibitem{soundararajan2012video}
Rajiv Soundararajan and Alan~C Bovik.
\newblock Video quality assessment by reduced reference spatio-temporal
  entropic differencing.
\newblock {\em IEEE Trans. Circuits Syst. Video Technol.}, 23(4):684--694,
  2012.

\bibitem{sun2022deep}
Wei Sun, Xiongkuo Min, Wei Lu, and Guangtao Zhai.
\newblock A deep learning based no-reference quality assessment model for ugc
  videos.
\newblock In {\em Proc. ACM Int. Conf. Multimed.}, pages 856--865, 2022.

\bibitem{thomee2016yfcc100m}
Bart Thomee, David~A Shamma, Gerald Friedland, Benjamin Elizalde, Karl Ni,
  Douglas Poland, Damian Borth, and Li-Jia Li.
\newblock Yfcc100m: The new data in multimedia research.
\newblock {\em Commun. ACM}, 59(2):64--73, 2016.

\bibitem{tu2021ugc}
Zhengzhong Tu, Yilin Wang, Neil Birkbeck, Balu Adsumilli, and Alan~C Bovik.
\newblock Ugc-vqa: Benchmarking blind video quality assessment for user
  generated content.
\newblock {\em IEEE Trans. Image Process.}, 30:4449--4464, 2021.

\bibitem{tu2021rapique}
Zhengzhong Tu, Xiangxu Yu, Yilin Wang, Neil Birkbeck, Balu Adsumilli, and
  Alan~C Bovik.
\newblock Rapique: Rapid and accurate video quality prediction of user
  generated content.
\newblock {\em IEEE Open J. Signal Process.}, 2:425--440, 2021.

\bibitem{vu2014vis}
Phong~V Vu and Damon~M Chandler.
\newblock Vi {S}3: an algorithm for video quality assessment via analysis of
  spatial and spatiotemporal slices.
\newblock {\em J. Electron. Imaging}, 23(1):013016--013016, 2014.

\bibitem{wang2005brightness}
Chao Wang and Zhongfu Ye.
\newblock Brightness preserving histogram equalization with maximum entropy: a
  variational perspective.
\newblock {\em IEEE Trans. Consum. Electron.}, 51(4):1326--1334, 2005.

\bibitem{wang2016mcl}
Haiqiang Wang, Weihao Gan, Sudeng Hu, Joe~Yuchieh Lin, Lina Jin, Longguang
  Song, Ping Wang, Ioannis Katsavounidis, Anne Aaron, and C-C~Jay Kuo.
\newblock {MCL-JCV}: a {JND}-based {H. 264/AVC} video quality assessment
  dataset.
\newblock In {\em Proc. IEEE Int. Conf. Image Process.}, pages 1509--1513,
  2016.

\bibitem{wang2018deep}
Miao Wang, Guo-Ye Yang, Jin-Kun Lin, Song-Hai Zhang, Ariel Shamir, Shao-Ping
  Lu, and Shi-Min Hu.
\newblock Deep online video stabilization with multi-grid warping
  transformation learning.
\newblock {\em IEEE Trans. Image Process.}, 28(5):2283--2292, 2018.

\bibitem{wang2019youtube}
Yilin Wang, Sasi Inguva, and Balu Adsumilli.
\newblock Youtube ugc dataset for video compression research.
\newblock In {\em Proc. IEEE Int. Workshop Multimed. Signal Process.}, pages
  1--5, 2019.

\bibitem{wang2004image}
Zhou Wang, Alan~C Bovik, Hamid~R Sheikh, and Eero~P Simoncelli.
\newblock Image quality assessment: from error visibility to structural
  similarity.
\newblock {\em IEEE Trans. Image Process.}, 13(4):600--612, 2004.

\bibitem{wu2022fast}
Haoning Wu, Chaofeng Chen, Jingwen Hou, Liang Liao, Annan Wang, Wenxiu Sun,
  Qiong Yan, and Weisi Lin.
\newblock Fast-vqa: Efficient end-to-end video quality assessment with fragment
  sampling.
\newblock In {\em Proc. Eur. Conf. Comput. Vis.}, pages 538--554, 2022.

\bibitem{yu2020learning}
Jiyang Yu and Ravi Ramamoorthi.
\newblock Learning video stabilization using optical flow.
\newblock In {\em Proc. IEEE Comput. Soc. Conf. Comput. Vision Pattern
  Recognit.}, pages 8159--8167, 2020.

\bibitem{zhang2022deep}
Zhao Zhang, Huan Zheng, Richang Hong, Mingliang Xu, Shuicheng Yan, and Meng
  Wang.
\newblock Deep color consistent network for low-light image enhancement.
\newblock In {\em Proc. IEEE Comput. Soc. Conf. Comput. Vision Pattern
  Recognit.}, pages 1899--1908, 2022.

\bibitem{zhao2020pwstablenet}
Minda Zhao and Qiang Ling.
\newblock Pwstablenet: Learning pixel-wise warping maps for video
  stabilization.
\newblock {\em IEEE Trans. Image Process.}, 29:3582--3595, 2020.

\bibitem{zheng2022semantic}
Shen Zheng and Gaurav Gupta.
\newblock Semantic-guided zero-shot learning for low-light image/video
  enhancement.
\newblock In {\em Proc. IEEE/CVF Winter Conf. Appl. Comput. Vis.}, pages
  581--590, 2022.

\bibitem{zhong2020efficient}
Zhihang Zhong, Ye Gao, Yinqiang Zheng, and Bo Zheng.
\newblock Efficient spatio-temporal recurrent neural network for video
  deblurring.
\newblock In {\em Proc. Eur. Conf. Comput. Vis.}, pages 191--207, 2020.

\end{thebibliography}
}

\end{document}